%% file: draft.tex
\documentclass[traditabstract]{aa}
\usepackage{graphicx}
\usepackage{txfonts}
\usepackage{natbib}
\usepackage{threeparttable}

\usepackage{color}

\newcommand{\ltsim}{\protect\raisebox{-0.5ex}{$\:\stackrel{\textstyle <}{\sim}\:$}}

\begin{document} 
  \title{Environmental dependence of polycyclic aromatic hydrocarbon emission at z$\sim$0.8
  }
  \subtitle{Investigation by observing the RX J0152.7-1357 with AKARI}
     
\author{Kazumi Murata
  \inst{1}
  \and Yusei Koyama\inst{1,2}
  \and Masayuki Tanaka\inst{2}
  \and Hideo Matsuhara\inst{1}
  \and Tadayuki Kodama\inst{2}
  %\fnmsep
  %\thanks{Just to show the usageof the elements in the author field}
}

\institute{
  Institute of Space and Astronautical Science, Japan Aerospace Exploration Agency, Sagamihara, Kanagawa 229-8510, Japan\\
  \email{murata@ir.isas.jaxa.jp}
  \and
  National Astronomical Observatory of Japan, Osawa 2-21-1, Mitaka, Tokyo 181-8588, Japan
}

\date{Received April 1, 2015; accepted June 29, 2015}
\abstract
    {We study the environmental dependence of the strength of polycyclic aromatic hydrocarbon (PAH) emission by AKARI observations of RX J0152.7-1357, a galaxy cluster at $z$=0.84.
      PAH emission reflects the physical conditions of galaxies and dominates 8 $\mu$m luminosity (L8), which can directly be measured with the L15 band of AKARI.
      $L8$ to infrared luminosity (LIR) ratio is used as a tracer of the PAH strength. 
      Both photometric and spectroscopic redshifts are applied to identify the cluster members.
      The L15-band-detected galaxies tend to reside in the outskirt of the cluster and have optically green colour, $R-z'\sim$ 1.2.
      We find no clear difference of the $L8$/LIR behaviour of galaxies in field and cluster environment. 
      The $L8$/LIR of cluster galaxies decreases with specific-star-formation rate divided by that of main-sequence galaxies, and with LIR, consistent with the results for field galaxies.
      The relation between $L8$/LIR and LIR is between those at $z=0$ and $z=2$ in the literature. 
      Our data also shows that starburst galaxies, which have lower $L8$/LIR than main-sequence, are located only in the outskirt of the cluster. 
      All these findings extend previous studies, indicating that environment affects only the fraction of galaxy types and does not affect the $L8$/LIR behaviour of star-forming galaxies.
    }
    \titlerunning {Environmental dependence of PAH emission at $z=0.8$}
    \maketitle

\input{intro}

\input{data}

\input{result}
\input{conc}
\appendix
\begin{acknowledgements}
This research is based on observations with AKARI, a JAXA project with the participation of ESA.
K.Murata is supported by JSPS (No.0263507).
This work was financially supported in part by a Grant-in-Aid for the Scientific Research (No.26800107).
\end{acknowledgements}

\bibliographystyle{aa}
\bibliography{rxj0152,pah}
\end{document}

%% file: intro.tex
\section{Introduction}
Galaxy formation and evolution are highly affected by its environment.
In the local Universe, galaxies in clusters tend to be red, early type \cite[]{1980ApJ...236..351D} and show lower star-forming activities. 
This star formation - density relation can also be seen in the distant universe \cite[]{2012ApJ...744...88Q}, while some studies suggest that the relation is reversed at $z\ge 1$ \cite[]{2007A&A...468...33E,2008MNRAS.383.1058C}.
Although star formation - density relation at higher redshift is still under debate, it is clear that the environment must play an important role in galaxy formation.
\par
Recent studies suggest that the environment affects only the fraction of galaxy types, and star-forming galaxies in different environment have similar properties \cite[]{2009ApJ...705L..67P,2012ApJ...746..188M,2013MNRAS.434..423K,2014ApJ...789...18K}.
\cite{2009ApJ...705L..67P} suggest that lower star formation rate (SFR) of galaxies in higher density may reflect the lower fraction of star forming galaxies, rather than a decrease of SFRs.
%Muzzin et al. shows that the fraction of star-forming galaxies is lower in cluster environment, while the specific SFR (sSFR) and the Dn(4000) of SF galaxies at fix stellar mass at $z\sim 1$ are independent of the environment.
%Koyama et al.(2013)
\cite{2013MNRAS.434..423K} discusses that SFR - stellar mass ($M_*$) relation of $\rm H\alpha$ selected galaxies is independent of the environment since $z\sim2$.
On the other hand, they also show that the dust attenuation and the median SFR of $\rm H\alpha$ emitters are higher in higher density environment at $z=0.4$, which can be seen only with infrared data.
Some studies reported that dusty star-forming galaxies reside in groups \cite[]{2008MNRAS.391.1758K,2011ApJ...734...66K,2009ApJ...705..809T}. 
These studies imply that dust has important roles in understanding the environmental effects on galaxy formation.
\par
%Among the dusts,
Polycyclic aromatic hydrocarbon (PAH) is dust showing prominent features at mid-infrared (MIR). 
They are excited by UV light from young stars and emit the energy at 3.3, 6.2, 7.7, 8.6 and 11.3 $\mu$m \cite[]{2007ApJ...657..810D}.
Among the PAH bands, 7.7 and 8.6 $\mu$m features dominate 8 $\mu$m luminosity ($L8$) even in broad band filters, so that $L8$ is used as a PAH luminosity tracer.
%Especially, 7.7 $\mu$m emission is the strongest and it dominates 8 $\mu$m luminosity ($L8$) even in broad band filters, so that $L8$ is used as a PAH luminosity tracer. 
Due to its energy sources, the PAH emission has been thought to correlate with the infrared luminosity (LIR), which is correlated with SFR.
Recent works suggest that high specific SFR (sSFR) galaxies tend to show weak PAH emission compared with main-sequence galaxies \cite[]{2011A&A...533A.119E,2012ApJ...745..182N,2014A&A...566A.136M}. 
%Elbaz et al. show that galaxies on the star-forming main-sequence has IR8 (LIR/$L8$) of $4\pm 1.6$.
The main causes of the PAH weakness is thought to be destruction of PAHs or a lack of UV photon that excites the PAHs. 
These studies indicate that the PAH emission traces the physical conditions of the interstellar matter rather than the SFR. 
On the other hand, the relation between infrared luminosity and PAH emission evolve with redshift \cite[]{2010A&A...514A...5T,2012ApJ...745..182N}. 
It suggests that the physical conditions of galaxies with given infrared luminosity is different at different redshift. 
Hence, if the cluster environment affects the galaxy properties, the relation between infrared luminosity and PAH emission may be different. 
\par
However, the behaviour of the PAH emission in cluster environments has not yet been studied well. 
%received little attention. 
This is mainly due to sparse filter sampling at 8-24 $\mu$m in the {\it Spitzer} space telescope \cite[]{2004ApJS..154....1W}, where 7.7 $\mu$m PAH emission of galaxies at $z<2$ is redshifted.
In contrast, Japanese AKARI satellite \cite[]{2007PASJ...59S.369M} has continuous wavelength coverage at 2-24 $\mu$m with nine photometric bands in the infrared camera (IRC; Onaka et al.2007).
It enables us to measure the PAH emission at $z<2$. 
%Using AKARI observation with the IRC, the behaviour of the PAH emission in cluster environments can be investigated.
\par
The galaxy cluster RX 0152.7-1357 (here after RXJ0152) at z$\sim$0.84 is one of the suitable targets for investigating an environmental dependence of dust properties.
It is one of the most distant clusters discovered by the {\it ROSAT} deep cluster survey \cite[]{1998ApJ...492L..21R}.
The $z\sim 0.8$ is suitable redshift, at which the rest frame 8 $\mu$m luminosity can directly be measured with the {\it AKARI}'s L15 band.
This cluster has been studied by many researchers, so that a number of ancillary data is available: spectroscopically confirmed members are provided in \cite{2005A&A...432..381D,2010ApJ...725.1252D,2006MNRAS.365.1392T,2005AJ....129.1249J}. 
It was observed by MIPS on-board {\it Spitzer} \cite[]{2007ApJ...654..825M}, by {\it Chandra} \cite[]{2007ApJS..168...19M}, by PACS and SPIRE on-board {\it Herschel} \cite[]{2011A&A...532A..90L,2012MNRAS.424.1614O} and by {\it Subaru}/Suprime-cam \cite[]{2005PASJ...57..309K}.
This cluster is known to have at least three X-ray clumps \cite[]{2005A&A...442...29G,2006ApJ...640..219M}, where two main clumps are expected to merge \cite[]{2005A&A...432..381D}. In these clump regions, most of galaxies are not star-forming \cite[]{2005ApJ...621..651H,2007ApJ...654..825M}, and exhibit early-type morphology \cite[]{2013A&A...555A...5N}. It is believed that they have formed the bulk of the stars in a short period burst at earlier epoch \cite[]{2014ApJ...797..136F}. 
\par
By observing the RXJ0152 with AKARI, we investigate the behaviour of the PAH emission of cluster galaxies, and compare it with the results for field galaxies presented in \cite{2014A&A...566A.136M} to complement the known cluster properties and to improve our understanding of environmental impact on galaxy evolution.
This work is organised as follows.
Data and methods are described in section 2.
Section 3 provides our results, and they are discussed and summarised in section 4.
Throughout this paper, we adopt a cosmology with ($\Omega_m$, $\Omega_\Lambda$, $H_0$) = (0.3, 0.7, 70 km s$^{-1}$ Mpc$^{-1}$), so that 1 arcsec corresponds to 7.6 kpc at $z=0.8$.
An initial mass function of \cite{2003PASP..115..763C} is assumed.

%% file: data.tex
\section{Data and Methods}
In this study, we investigate PAH behaviour of RXJ0152 galaxies and compare it with that of field galaxies studied in \cite{2014A&A...566A.136M}.
We adopt $L8$ to LIR ratio as a tracer of PAH strength, and ratio of LIR to $M_*$ as starburst strength, following our previous studies on the AKARI north ecliptic pole deep survey \cite[hereafter NEP-Deep]{2006PASJ...58..673M,2008PASJ...60S.517W,2013A&A...559A.132M}.
We derive these quantities using {\it AKARI}/IRC, {\it Spitzer}/IRAC and MIPS, {\it Herschel}/PACS and SPIRE, and {\it Subaru}/S-cam.

\subsection{{\it AKARI}/IRC}
Twelve {\it AKARI}/IRC pointed observations were conducted on 2007 July 12-13 using S7 and L15 bands with AOT05 astronomical templates \cite[]{2007PASJ...59S.401O}.
The AOT05 takes $\sim$30 sub-images with 16.4 seconds without dithering and filter change to obtain long exposure time.
%The S7 and the L15 band images, whose field of views are separated by 20 arcmin, are taken simultaneously so that the alignment correction should be the same.
The field of view of the S7 and L15 band images are separated by $\sim$20 arcmin. 
The observations consist of two fields, F1(RA=28.1679,DEC=-13.9683) and F2 (RA=28.3029,DEC=-13.9141).
The F1 field was observed four times with the L15 band and three times with the S7, while the F2 field was observed twice with the L15 and three times with the S7.
\par
\begin{figure}
  \centering
  \includegraphics[width=88mm]{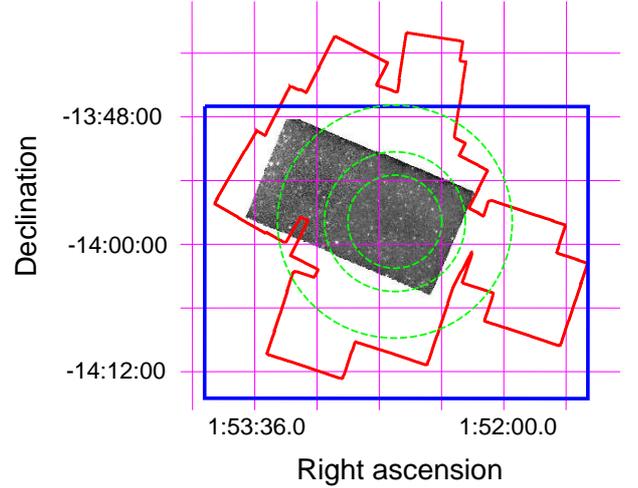}
  \caption{Mosaicked L15 image ($\sim 10 \times 20$ arcmin$^2$).
    The blue square and the red line indicate {\it Subaru}/S-cam and {\it Spitzer}/MIPS coverages. 
    The green broken circles indicate 2, 3, and 5 Mpc radii from the cluster centre.
%    The magenta and cyan points indicate the cluster member galaxies with photometric and spectroscopic redshifts, respectively.
  \label{fig:L15map}}
\end{figure}

The image reduction was performed on the basis of standard manner.
The linearity correction, the dark subtraction, and the flat fielding were done with the {\it AKARI}/IRC imaging pipeline (version 131202).
%Dark frame taken in each observation was used for the dark subtraction.
The other detail reduction was performed in the same way in \cite{2013A&A...559A.132M}.
The world coordinate system (WCS) information was registered using {\it Spitzer}/IRAC and MIPS images downloaded from the IRSA web site\footnote{http://irsa.ipac.caltech.edu/Missions/spitzer.html}.
Bad pixels were removed using the all-dark-frame-stacked images. 
The sky background was subtracted from each sub-frame using a sky map provided by SExtractor \cite[]{1996A&AS..117..393B}. 
Finally, all images were mosaicked, during which the pixel sizes were reduced to 1$\times$1 arcsec$^2$ from 2.34$\times$2.34 and 2.51$\times$2.39 arcsec$^2$ for the S7 and L15 band images, respectively. 
Figure \ref{fig:L15map} shows the mosaicked L15 image, which covers a $\sim$ 10 $\times$ 20 arcmin$^2$ region. 
The green broken circles indicates 2, 3, and 5 Mpc from the cluster centre, RA=28.175, DEC=-13.965 \cite[]{2000ApJS..126..209R}.
The L15 image covers the region of $R<2$ Mpc around the cluster. 

\par 
Source extraction was performed on the mosaic images using the SExtractor.
Sources with five connected pixels having above 1$\sigma$ from the local background were detected.
%Although this detection criteria may be too generous, which leads some fake sources, they were removed from the final sample when matching with other catalogues.
Although this detection criteria may be too generous, which leads some fake sources, they were removed from the final sample because we use only objects with an optical and MIPS counterpart. 
%when matching with other catalogues.
The source fluxes were measured with 6 arcsec aperture radius, for which \cite{2013A&A...559A.132M} estimated the photometric calibration.
The 5$\sigma$ detection limits estimated with random sky photometry were 76 and 268 $\mu$Jy for the S7 and L15 bands, respectively.
These values may be overestimated due to residual sky background. 
The flux errors derived with the sky deviation in the annulus of 20 arcsec with 15 arcsec width were scaled to be consistent with the above detection limits. 
Although it may lead overestimation, we decided to apply these conservative errors.

\subsection{{\it Spitzer}/IRAC and MIPS}
We retrieved the {\it Spitzer} data from the IRSA website.
All post-basic-calibrated data (PBCD) around the cluster was downloaded for both IRAC and MIPS. 
Sky background was subtracted from each image using a median filtered image after masking the sources.
After the sky subtraction, all images were aligned and combined with median values. 
The red line in Fig.\ref{fig:L15map} shows the coverage map of the MIPS image, which covers $R<$ 3 Mpc from the cluster centre.
The IRAC image also covers $R\ltsim$ 3 Mpc, although it is not shown in the figure for simplicity.
%\color{red}{largaer than marcillac. typical exposure time 2ks. deviation?}
%\color{black}
\par
Source extraction was performed using the SExtractor. 
Sources with five connected pixels above 1$\sigma$ from the background were extracted. 
Similar to {\it AKARI} sources, fake sources were contained, but they can be removed because we use only objects with an optical counterpart. 
%when matching with other catalogues. 
Photometry was simultaneously done for the IRAC 1-4 channels with the dual mode of the SExtractor, where we used the [3.6] band image as the detection image and the aperture radius was set to 6 arcsec.
Photometry for the MIPS 24 $\mu$m band was done with a 7 arcsec aperture radius and the sky background was estimated from the annulus of 15 arcsec with 5 arcsec width.
While the sky background in the IRAC images was subtracted by the SExtractor, we determined the sky background level of MIPS image with an annulus around each source. This is because the MIPS image has some artificial structures and the sky background should be measured from a region near the source position. 
The flux errors were estimated in the same way as we did for the {\it AKARI} sources. 

\begin{figure}
  \includegraphics[angle=270,width=80mm]{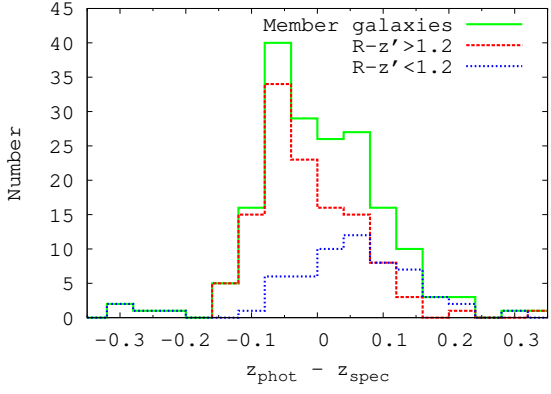}
  \caption{Distribution of $z_{phot}-z_{spec}$ for 186 member galaxies.
    The red and blue histograms indicate the number of optically red ($R-z'>1.2$) and blue ($R-z'<1.2$) galaxies, respectively. 
%  The standard deviation is $\sim$0.08.
%  Among 186 galaxies, 10 members have outlier values (3$\sigma$) 
  \label{fig:zpzs}}
\end{figure}

\subsection{{\it Herschel}/PACS and SPIRE}
{\it Herschel} observed the RXJ0152 as PACS evolutionary probe (PEP; Lutz et al.2011), and the catalogue of 100 and 160 $\mu$m bands of the PACS and 250, 350, and 500 $\mu$m bands of the SPIRE are available. 
We downloaded all these catalogues from the CDS\footnote{http://cdsweb.u-strasbg.fr/}. 
Although it is also observed with SPIRE as {\it Herschel} multi-tiered extragalactic survey \cite[]{2012MNRAS.424.1614O}, we did not use these data for simplicity.

\subsection{{\it Subaru}/Suprime-Cam}
{\it Subaru}/S-cam $V R i' z'$ images were obtained in \cite{2005PASJ...57..309K} and \cite{2005MNRAS.362..268T}.
The reduction procedure is described in \cite{2005PASJ...57..309K}.
The detection limits of these bands with 2 arcsec aperture were reported as 26.7, 26.5, 26.1, and 25.0, for $V$, $R$, $i'$, and $z'$ bands, respectively.
The coverage is shown in Fig.\ref{fig:L15map} with the blue square.
Source extraction was also done using the SExtractor and objects with 5 connected pixels whose values were more than 1.5$\sigma$ above the background were detected.
These detection criteria are the default parameters of the SExtractor.
Photometry was performed with the dual mode using the $z'$ band image as the detection image, and $MAG\_AUTO$ was applied for measuring the total magnitudes.
Because magnitude errors output from the SExtractor is underestimated, we added 0.05 magnitude to the errors, although it is arbitrary. 
%We added 0.05 magnitude errors, which are arbitrary, for uncertainty of magnitude zero points for all bands. 

\begin{figure}
  \includegraphics[width=60mm,angle=270]{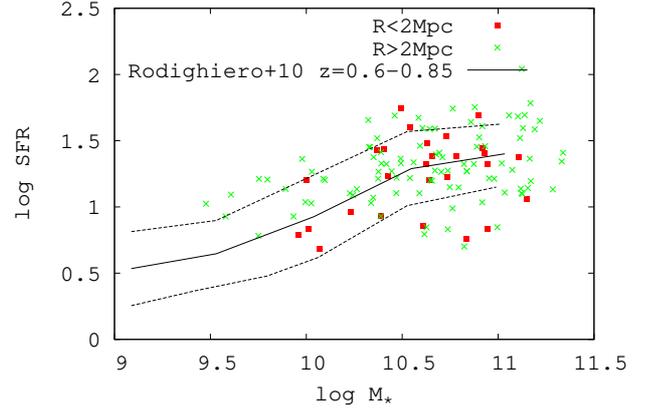}
  \caption{Star-formation rate versus stellar mass.
    The red squares and the green crosses indicate galaxies located in $R <$ 2 Mpc and $R>$ 2 Mpc, respectively. 
    The black solid and broken lines indicate the average and the 1$\sigma$ range of main-sequence relation for field galaxies at $z=$0.6-0.85 \cite[]{2010A&A...518L..25R}.
  \label{fig:ms}}
\end{figure}

\subsection{Matching the catalogues}
To merge the above catalogues, the sources were cross-matched with the nearest neighbour method. 
{\it Subaru} sources were matched with {\it Spitzer}/IRAC sources with a 1 arcsec search radius.
IRAC sources were matched with AKARI/S7, L15, and MIPS24 sources with 2.5 arcsec radii, respectively, where the size of the 2.5 arcsec search radius is the same as the matching radius used in \cite{2013A&A...559A.132M}. 
Finally, {\it Herschel}/PACS and SPIRE sources were matched with MIPS24 sources with 5 and 6 arcsec radii, which are the same as in \cite{2014A&A...566A.136M}.
We used only objects that have both IRAC and {\it Subaru} photometry to avoid fake sources and to make photometric redshifts reliable.
Also, L15-detected objects without MIPS 24 $\mu$m photometry, which may be due to matching errors, was not used in this study. 
\par
We also matched these sources with spectroscopically confirmed members provided by previous studies \cite[]{2005A&A...432..381D,2010ApJ...725.1252D,2005AJ....129.1249J,2006ApJ...644...30B,2006MNRAS.365.1392T}.
All these sources were matched with {\it Subaru} sources with a 1 arcsec search radius.
The redshift of the member galaxies from \cite{2006ApJ...644...30B} were assumed to be $z$=0.84.
Following \cite{2009ApJ...694.1349P}, we assumed objects with $z_{spec}$=0.80-0.87 are member galaxies.
A total of 186 spectroscopically confirmed members were matched with our catalogue, among which 17 objects were detected with both the {\it AKARI} L15 band and the MIPS24 band.
The brightest cluster galaxies defined in \cite{2010ApJ...718...23S} and \cite{2013MNRAS.433..825L} are not detected with the L15 band.
We note that 41 member objects were rejected due to a lack of IRAC photometry. 
We also note that X-ray sources from \cite{2007ApJS..168...19M} are not matched with our catalogue due to blending other sources, although it can visually be confirmed. 
Hence, we assume that a strong active galactic nucleus is not in our sample.

\subsection{Photometric redshifts}
The photometric redshifts were calculated using a publicly available software, LePHARE \cite[]{2006A&A...457..841I,2007A&A...476..137A}.
It performs a template fitting with the spectral energy distribution (SED) for each object, and calculates photometric redshift as well as physical parameters, with which the minimum $\chi^2$ is provided.
%to minimise the $\chi^2$.
We used AVEROIN SED templates \cite[]{2007A&A...476..137A} with an interstellar extinction law of SMC following \cite{2007A&A...476..137A}.
Following Murata et al.(2014), the AVEROIN templates were used only for estimating photometric redshifts because the AVEROIN is suitable for photometric redshift estimation due to small parameter sets. 
The {\it Subaru}/S-cam $VRi'z'$ to IRAC [3.6] and [4.5] were fitted with the templates, where [3.6] and [4.5] covers longer wavelength than a stellar bump at 1.6 $\mu$m. 
%We applied the ``AUTO\_ADOPT'' option, which calculates photometric zero point correction to minimise the difference between SEDs of objects with spectroscopic redshift and best fit templates. 
\par
The accuracy of the photometric redshifts was calculated with spectroscopic redshifts. 
Figure \ref{fig:zpzs} shows the distribution of the difference of photometric and spectroscopic redshifts for the 186 member galaxies.
The histograms for blue and red galaxies are also shown for comparison.
Although the averages of the $z_{phot}-z_{spec}$ for red and blue galaxies have an offset from zero, the average of all sample without outliers is $z_{phot}-z_{spec}$ = -0.0038.
The dependence of the accuracy on optical colours is also reported in Tanaka et al.(2006). Although their sample shows that bluer galaxies have lower photometric redshift, our sample showed an opposite trend. It can be due to the difference of SED templates used in the estimation. The colour dependence of the photometric redshift accuracy affects the fraction of star-forming galaxies, as Tanaka et al. discussed. Nonetheless, this effect should be similar in both the cluster core and the outskirt and our conclusion should not be affected by this effect.
%It can be due to the extinction law, because the extinction can be overestimated due to the step size in the fitting, which leads to overestimate the photometric redshift.
The standard deviation is $\sigma_{zp}$ = 0.080, which corresponds to $\delta z_{phot}/(1+z_{spec})$=0.044.
Ten outliers having $z_{phot}>3\sigma$ were clipped in the calculation. 
The $z_{phot}$ average and standard deviation for the member galaxies are $z_{phot}$=0.843$\pm$0.0825, and therefore we assumed that objects with $z_{phot}$=0.76-0.925 are cluster member galaxies.
Although a fraction of member galaxies are missed in the criteria, we decided to apply it to avoid a increase of interlopers.
\par
In total, 1922 galaxies were identified as member galaxies, among which 122 have  MIPS photometry.
Among the total sample, 333 galaxies were located in the $R<$ 2 Mpc region, of which 38 objects were detected with the {\it AKARI} L15 band. 
%The locations of member galaxies are shown in Fig.\ref{fig:L15map}. 
%We can see a clear concentration on the cluster centre. 
We visually checked the SEDs of these MIPS detected sources, and found no galaxies with an unusual SED. 
In the following, the redshift of cluster members without spectroscopic redshift is set to $z=0.84$. 
\begin{figure*}[ht]
  \centering
  \begin{minipage}{0.49\hsize}
    \includegraphics[width=50mm,angle=270]{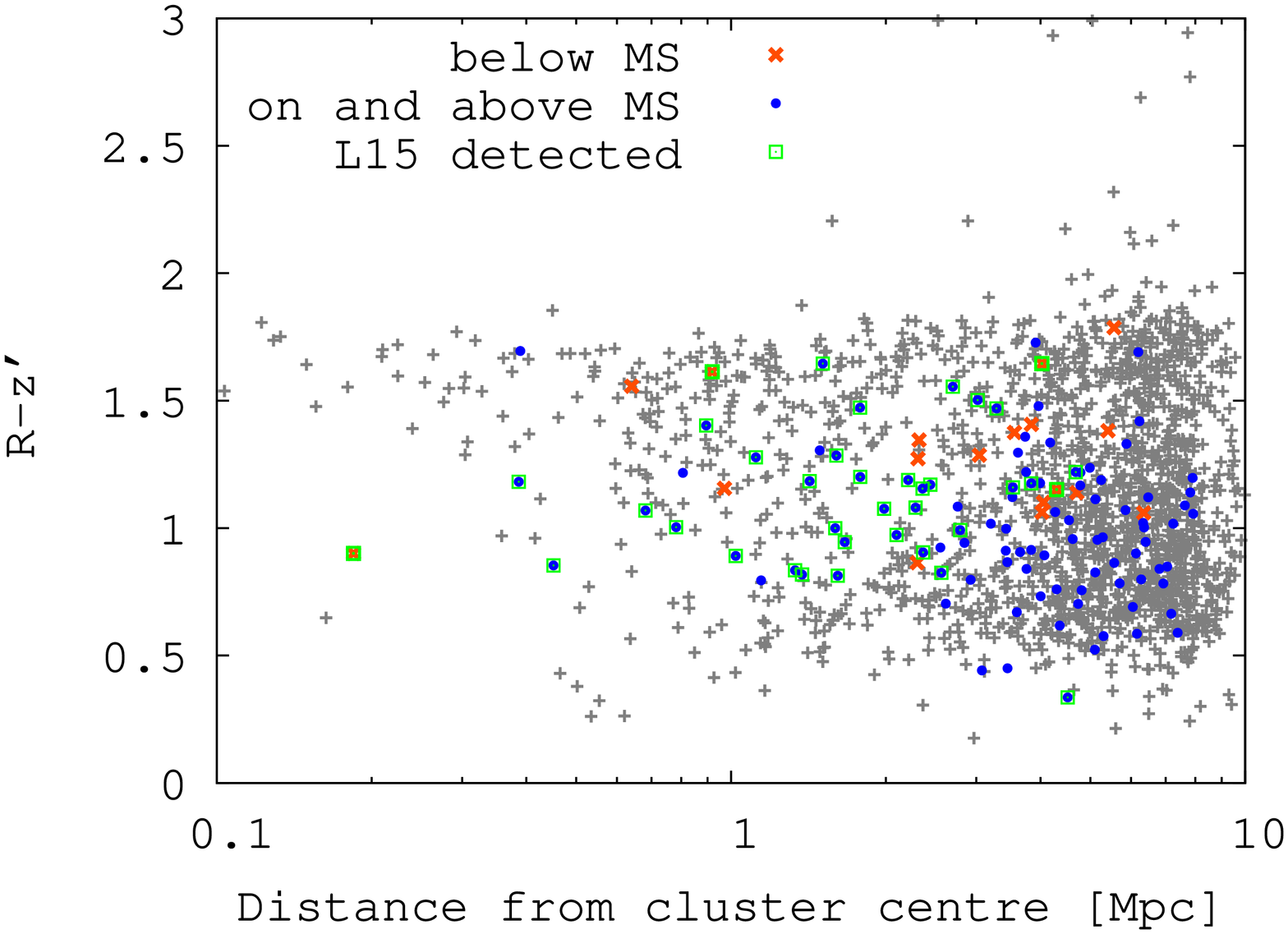}
  \end{minipage}
  \begin{minipage}{0.49\hsize}
    \includegraphics[width=50mm,angle=270]{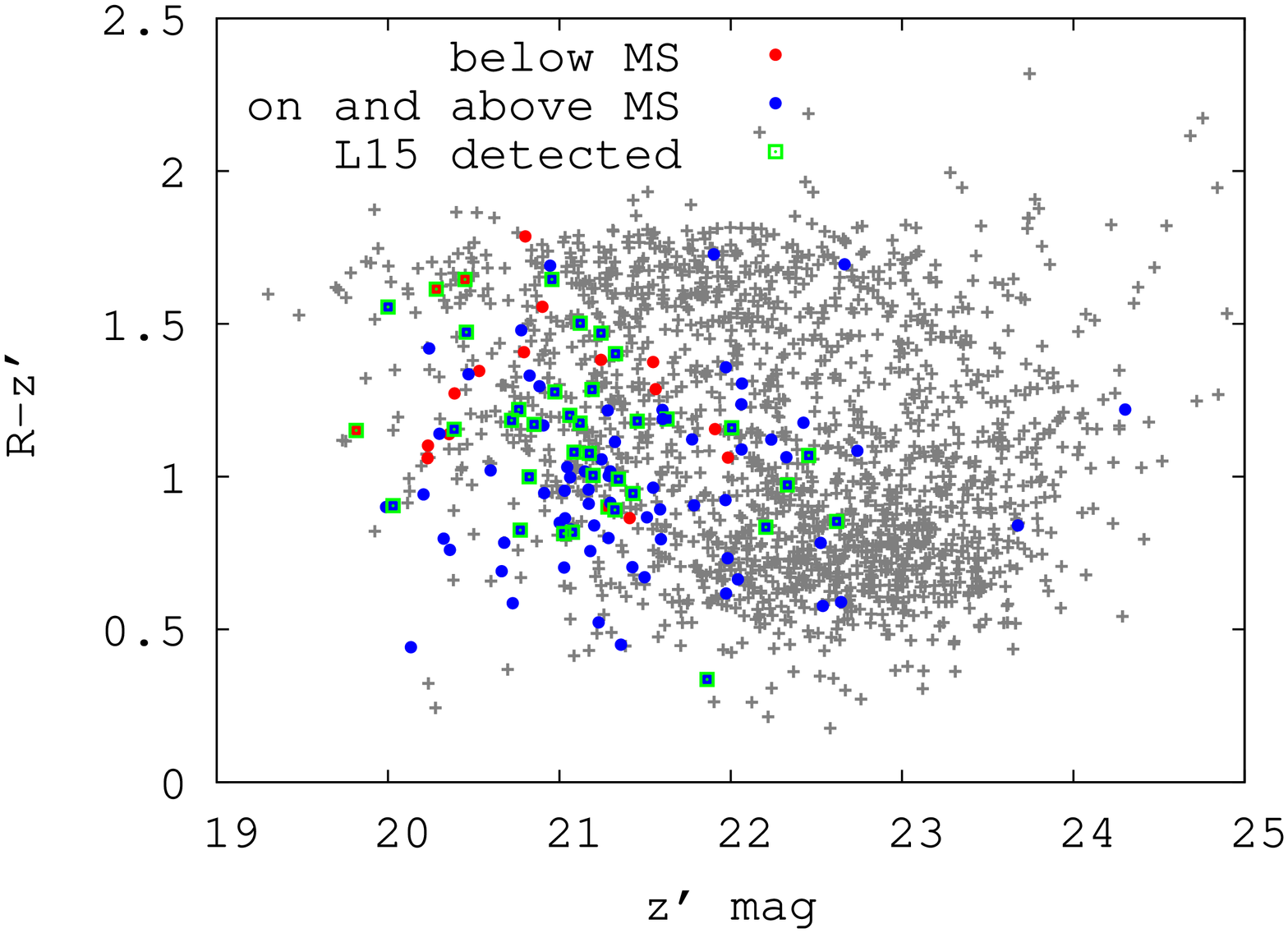}
  \end{minipage}
  \caption{
    Left: $R-z'$ colour vs distance from the cluster centre.
    The red crosses and the blue circles indicate below and on/above main-sequence galaxies, classified in Fig.\ref{fig:ms}.
%    passive and star-forming galaxies detected with MIPS24.   
%    see Fig.\ref{fig:ms} for the classification.              
    L15 detected galaxies are indicated by the green squares.
    MIPS24 non-detected galaxies are also plotted with the grey points.
    Right: $R-z'$ vs $z'$ colour-magnitude diagram.
    The symbols are the same as the left panel.
    \label{fig:colour}}
\end{figure*}

\subsection{Deriving physical parameters with SED fitting}
The main purpose of this paper is to compare the $L8$/LIR behaviour of cluster and field galaxies at $z=0.8$. 
We derive LIR, $M_*$, and $L8$ with the LePHARE in the same way as \cite{2014A&A...566A.136M}. 
\par
Infrared luminosity, LIR, integrated over 8 $\mu$m to 1000 $\mu$m, was estimated with the 24 $\mu$m to 500 $\mu$m photometry.
We did not use the L15 photometry because it was used to derive $L8$ and to avoid any dependency between $L8$ and LIR. 
An SED library of \cite{2001ApJ...556..562C} was used for the calculation.
%For a total of 122 galaxies, among which 22 have $Herschel$ photometry, LIR was derived.
For a total of 122 galaxies LIR was derived, among which 30 have $Herschel$ photometry. 
%magerr<1. >1を含めると145
Although for most of galaxies the LIR were derived with only the 24 $\mu$m luminosity, they can be reliable since \cite{2010A&A...518L..29E} show infrared luminosity derived from only 24 $\mu$m band agrees with that from 24 $\mu$m and 100-500 $\mu$m bands within 0.15 dex for galaxies at $z<1.5$.
The LIR distribution in our sample has a peak at log LIR $\sim$ 11.2, similar to the NEP-Deep field at the same redshift range.% However, it indicates that fainter galaxies with log LIR $\sim$ 11.2 have lower completeness, which can affect the results. This is also discussed in section \ref{l8}.}

\par
Stellar mass, $M_*$, was derived with $VRi'z'$, [3.6] and [4.5] photometry using an SED library of BC03 \cite[]{2003MNRAS.344.1000B}.
Since these six bands bracket 4000 $\AA$ breaks, $M_*$ can be reliably determined. 
We adopted SEDs of solar metallicity with $\tau$ = 0.1-10 Gyr and an extinction law of \cite{2000ApJ...533..682C}.
%The $AUTO\_ADOPT$ option was also applied.
We note again that the libraries used for stellar mass, photometric redshift, and infrared luminosity are different, because stellar mass estimation need more parameters than photometric redshift estimation and because dominant components for stellar mass and infrared luminosity are different so that they should be independently determined. 
To check whether we can see the star-formation main sequence, we plot the star-formation rate, which is the infrared luminosity multiplied by 1.09$\times$10$^{-10}$[SFR/L$_{\odot}$], and the stellar mass in Fig.\ref{fig:ms}. 
The star-formation main-sequence at $z=$ 0.6-0.85 from \cite{2010A&A...518L..25R} is also shown, where the difference of the initial mass function was corrected. 
We can see that both galaxies at $R<2$ Mpc (red squares) and at $R>2$ Mpc (green crosses) are on the same relation.
Some galaxies are above and below the main-sequence range.
Hereafter we call these galaxies galaxies above and below main-sequence. 
\par
$L8$ was derived with less uncertainty from $K$-correction using the L15 flux.
This is because the 8 $\mu$m band is redshifted into the L15 band at $z\sim0.8$. 
The $K$-correction was estimated with 25 SED templates from \cite{2006ApJ...642..673P,2007ApJ...663...81P}.
The typical value of the $K$-correction is only $\sim$0.2 dex.
The IRAC4 band was used for the 8 $\mu$m band for the $K$-correction.
Nearly 80\% of our sample have log $L8/\rm L_\odot > 10.3$.
We note that, although the L15 band almost matches the rest frame IRAC4 band, a fraction of silicate absorption at 9.7 $\mu$m could affect the $L8$ luminosity. Although this effect should be corrected by the $K$-correction for most of our sample, we estimated how $L8$ is affected in the worst case. We used an Arp220 template, which shows a very strong silicate absorption, and artificially masked the silicate feature to compare the L15 flux with and without the silicate feature. We confirm that 0.2 dex of the L15 flux can be reduced by the silicate absorption. However, even in the worst case, decrease of $L8$ in our results ($\sim$0.4 dex) cannot be explained by the silicate feature and the comparison with previous studies that derive $L8$ with similar methods is reliable.

%% file: result.tex
\section{Results}
Previous studies suggest that when only star-forming galaxies are considered, the local environment does not affect galaxy properties.   
Our purpose is to extend the previous studies and to investigate whether $L8$, which is dominated by PAH emission, is dependent on the environment.
%, since it traces the physical conditions of galaxies rather than the SFR.
To measure the $L8$ with less uncertainty of $K$-correction, we used the L15 band photometry.
Here we provide, in section \ref{colour}, the location (i.e. the environment) and the optical colour of galaxies to know the properties of galaxies detected with the L15 band in detail.
The location of starburst galaxies with respect to the cluster core is shown in section \ref{l8loca}.
Then, in section \ref{l8}, we investigate the $L8$/LIR behaviour and compare it with those of field galaxies.

\subsection{Colour and location of infrared galaxies}
\label{colour}
At $z\sim0.8$, \cite{2008MNRAS.391.1758K} reported that few star-forming galaxies are detected in cluster cores, and tend to be located in the medium-density region, at which optical colour shows a dramatic change. 
They also show that some 15 $\mu$m detected galaxies in RXJ1716.4+6708, a galaxy cluster at $z=0.81$, have optically red colours with $R$ - $z'$ $\ge$ 1.5, which are candidates of dusty red galaxies. 
\cite{2007ApJ...654..825M} show that the 24 $\mu$m detected galaxies in the RXJ0152, the same cluster as this study, tend to lie in the outskirts, and that they have similar colours to the late type member galaxies without 24 $\mu$m detection.
In our study, we show optical colours of on/above and below the main-sequence galaxies, compared with those of infrared undetected galaxies in the RXJ0152. 
\par
The left panel of Fig.\ref{fig:colour} shows the $R$-$z'$ colour against distance from the cluster centre.
Most of the galaxies in the $R < 1$ Mpc region have red $R-z'$ colour while outer galaxies show a bimodal colour distribution, consistent with \cite{2009ApJ...694.1349P}.
On the other hand, few IR galaxies are located in $R<1$ Mpc and a fraction of IR galaxies located at $R\sim1$ Mpc, which is consistent with \cite{2007ApJ...654..825M}. 
The galaxies at $R\sim1$ Mpc show a green colour, $R-z'\sim1.2$.
Considering this region is 2-3 times denser than the average so that inter mediate density region, calculated with 5th neighbour galaxies, it is consistent with previous studies. 
Galaxies below the main-sequence (red crosses) has redder colour than galaxies on/above the main-sequence (blue circles).
Galaxies on and above the main-sequence are merged in Fig.\ref{fig:colour} for simplicity.
The average colour of below, on and above galaxies are
$R-z'=$ 1.20 $\pm$ 0.05, 1.04 $\pm$ 0.03, and 0.94 $\pm$ 0.05, respectively.
Considering dusty star-formation is strongest in galaxies above the main-sequence and weakest in galaxies below the main-sequence, the above results indicate that the optical colours are more sensitive to the star-forming activity than the dust extinction. 
IR galaxies at $R>3$ Mpc seem to have bluer colour than those at $R\sim1$ Mpc, which is consistent with \cite{2009ApJ...705L..67P} who show that the sSFR derived from the MIPS 24 $\mu$m band decreases with increasing local density.
We visually checked the 2 dimension image that these galaxies were randomly distributed so that they are not likely affected by a special clump or any instrumental error. 
\par
The right panel of Fig.\ref{fig:colour} shows the $R-z'$ versus $z'$ colour magnitude diagram for below and on/above the main-sequence galaxies, L15 detected and IR undetected galaxies.
The L15-detected galaxies show, again, a green colour, which is consistent with \cite{2008MNRAS.391.1758K}. 
Most of L15-detected galaxies are brighter than $z'$=22 mag, due to the detection limits. 
Some L15-detected galaxies show a red colour, $R-z'>$1.5, which is also consistent with \cite{2008MNRAS.391.1758K}.
%, who discuss these galaxies are candidates of dusty star-forming galaxies. 
Although the L15-detected galaxies seem to be redder than other IR galaxies, it is a selection effect due to the limited coverage of the L15 image (see Fig.\ref{fig:L15map}) as can be seen in the left panel of Fig.\ref{fig:colour}.
%and IR galaxies in this region are redder than in the outer region,
%that these galaxies locate at $R<3$ Mpc due to the coverage

\subsection{Location of starburst galaxies}
\label{l8loca}
\begin{figure*}[t]
  \centering
  \includegraphics[height=100mm,angle=270]{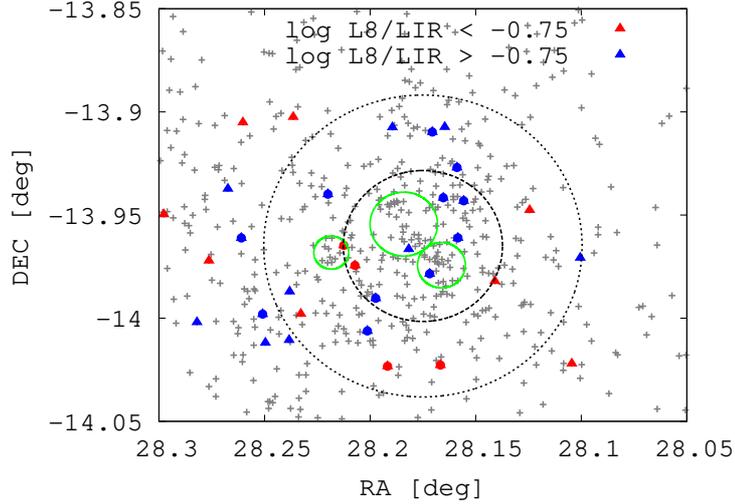}
  \caption{Location of L15 detected galaxies. The circles and the triangles indicate galaxies with and without a spectroscopic redshift, respectively. The red and blue symbols indicate those with log $L8$/LIR $<$ -0.75 and log $L8$/LIR $>$ -0.75, respectively. This boundary corresponds to 1$\sigma$ smaller than that of main-sequence galaxies determined in \cite{2011A&A...533A.119E}. The grey crosses indicate the member galaxies without L15 detection. The Black dotted lines indicate the 1 and 2 Mpc radii from the cluster centre. The green solid lines indicate the three clumps in \cite{2005A&A...432..381D}.  \label{fig:spdist}}
\end{figure*}
Here we investigate the location of starburst galaxies with respect to the cluster cores. In Fig.\ref{fig:spdist}, the galaxies are divided with $L8$/LIR. Galaxies with log $L8$/LIR $<$ -0.75, which corresponds to 1$\sigma$ smaller than that of main-sequence galaxies \cite[]{2011A&A...533A.119E}, are regarded as starburst galaxies. The green circles in Fig.\ref{fig:spdist} indicate the three clumps where above 3$\sigma$ X-ray emission is encompassed \cite[]{2005A&A...432..381D}. We note that the $R > 2 $ Mpc region is covered only partially with the L15 image (see Fig.\ref{fig:L15map}).

\par
\begin{figure*}
  \centering
  \begin{minipage}{0.49\hsize}
    \includegraphics[height=70mm,angle=270]{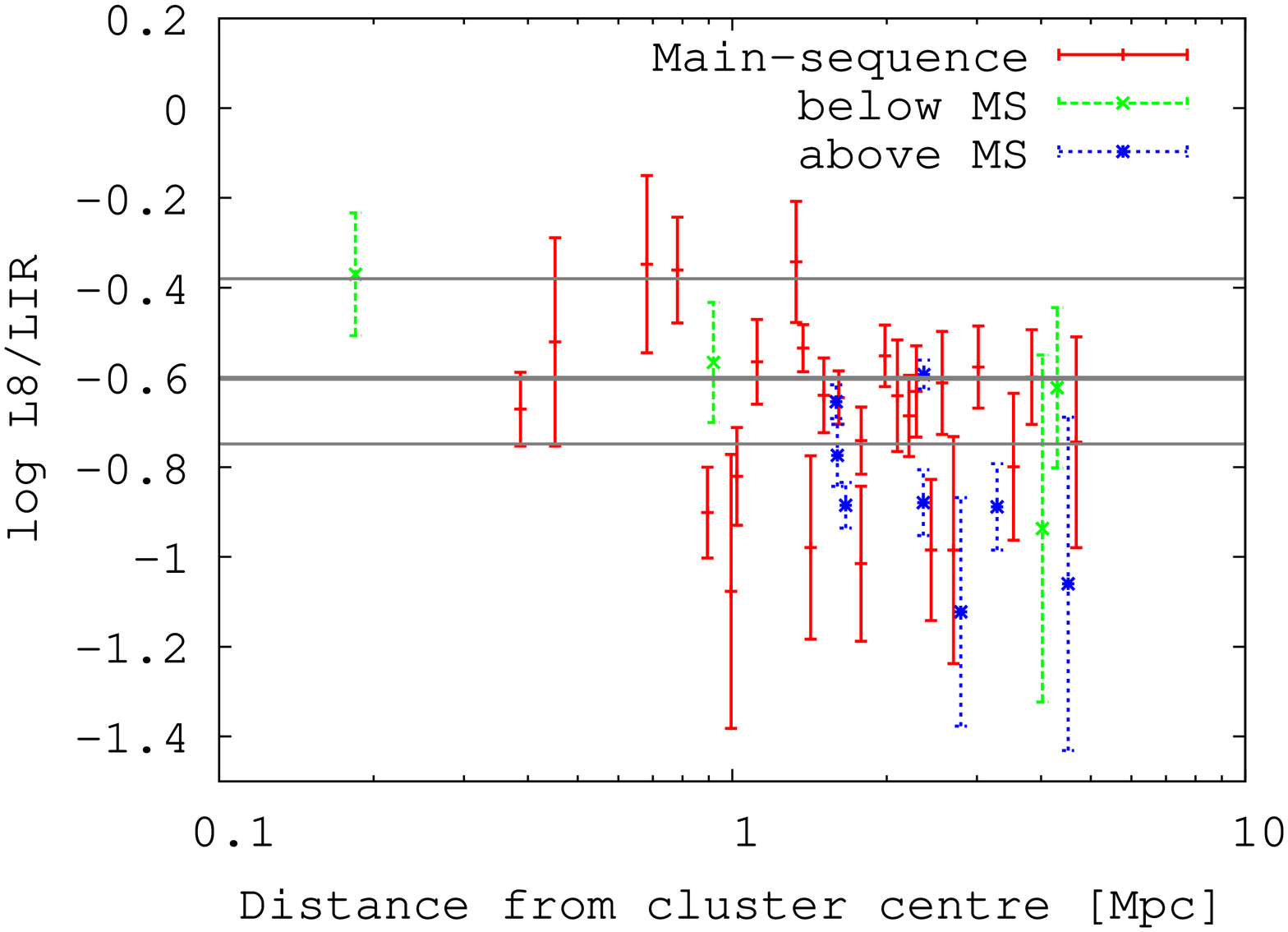}
  \end{minipage}
  \begin{minipage}{0.49\hsize}
    \includegraphics[height=70mm,angle=270]{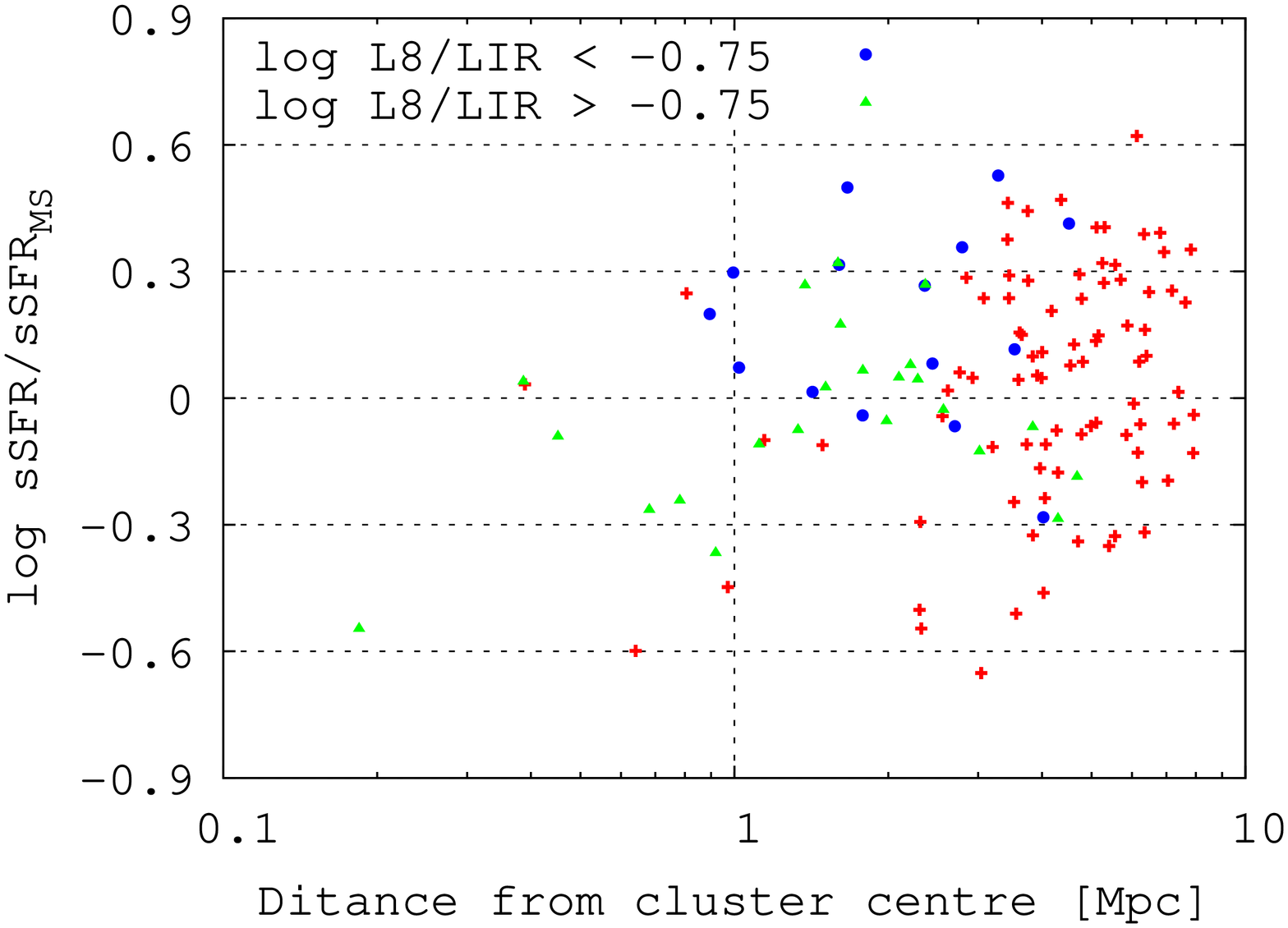}
  \end{minipage}
  \caption{
    Left:$L8$/LIR vs distance from the cluster centre. 
    The green, red, and blue points indicate the below, on, and above main-sequence galaxies in Fig.\ref{fig:ms}.
    The grey lines indicate the average and the range of $L8$/LIR for main-sequence galaxies in \cite{2011A&A...533A.119E}.
    Right: sSFR divided by that of main-sequence from \cite{2010A&A...518L..25R}.
    The blue circles and the green triangles indicate galaxies with log $L8$/LIR $<$ -0.75 and $>$ -0.75.
    The red crosses indicates galaxies without L15 band detection. 
    \label{fig:dis_l8lir}}
\end{figure*}

We can see that few galaxies are located in the clump, consistent with previous studies, who show a lack of star-forming galaxies in the cluster centre and X-ray clumps \cite[]{2005ApJ...621..651H,2007ApJ...654..825M}. Fig.\ref{fig:spdist} also shows that low $L8$/LIR galaxies tend to be located in the outskirt of the cluster. As \cite{2014ApJ...797..136F} discuss that galaxies in the outskirt show a variety of star formation histories, our sample also shows that both low and high $L8$/LIR galaxies reside in the outskirt. This can also be seen in Fig.\ref{fig:dis_l8lir}. 
In the left panel, the $L8$/LIR is plotted against the distance from the cluster centre.
We have to note that the error bars only indicate the $L8$ uncertainty.
Galaxies below, on, and above the main-sequence are indicated with the green, red, and blue points for a comparison. 
The average and the 1$\sigma$ range of the $L8$/LIR for main-sequence galaxies in \cite{2011A&A...533A.119E} are indicated with grey lines.
We can see that galaxies with log $L8$/LIR $<$ -0.75 are only seen at $R>1$ Mpc regions, while higher $L8$/LIR galaxies are located in the entire field.
Similarly, we show in the right panel that galaxies with log sSFR/sSFR$\rm _{MS}$ above the 0.3 dex scatter (see Fig.\ref{fig:ms}) are located only at $R>1$ Mpc. 
These results imply that galaxies in a cluster core at $z\sim0.8$ are not in a starburst mode.

\subsection{Behaviour of $L8$/LIR against LIR and LIR/$M_*$}
\label{l8}
Previous studies for field galaxies show $L8$ correlate with LIR at lower sSFR, but have a relative weakness of $L8$ at higher sSFR, throughout a redshift range of $z$=0.3-2 \cite[]{2011A&A...533A.119E,2012ApJ...745..182N,2014A&A...566A.136M}.
In our study, we investigate whether the $L8$/LIR behaviour of cluster galaxies is different from those of the field galaxies using the {\it AKARI} L15 photometry. 
Our results are compared with those of {\it AKARI} NEP-Deep field studied in \cite{2014A&A...566A.136M}, who applied similar methods to our work. 
We re-calculated their result from the same data set to match the redshift range with the current sample, $z_{phot}$ = 0.76-0.925. 
%Although they apply the {\it starburstiness}, offset of sSFR from that of the main-sequence, we use $LIR/M_*$ instead, because the offset only corrects the redshift dependence \cite[]{2011A&A...533A.119E} and our study focuses only on galaxies at $z\sim0.8$. 
\par
\begin{figure*}[ht]
  \centering
  \begin{minipage}{0.49\hsize}
    \includegraphics[width=50mm,angle=270]{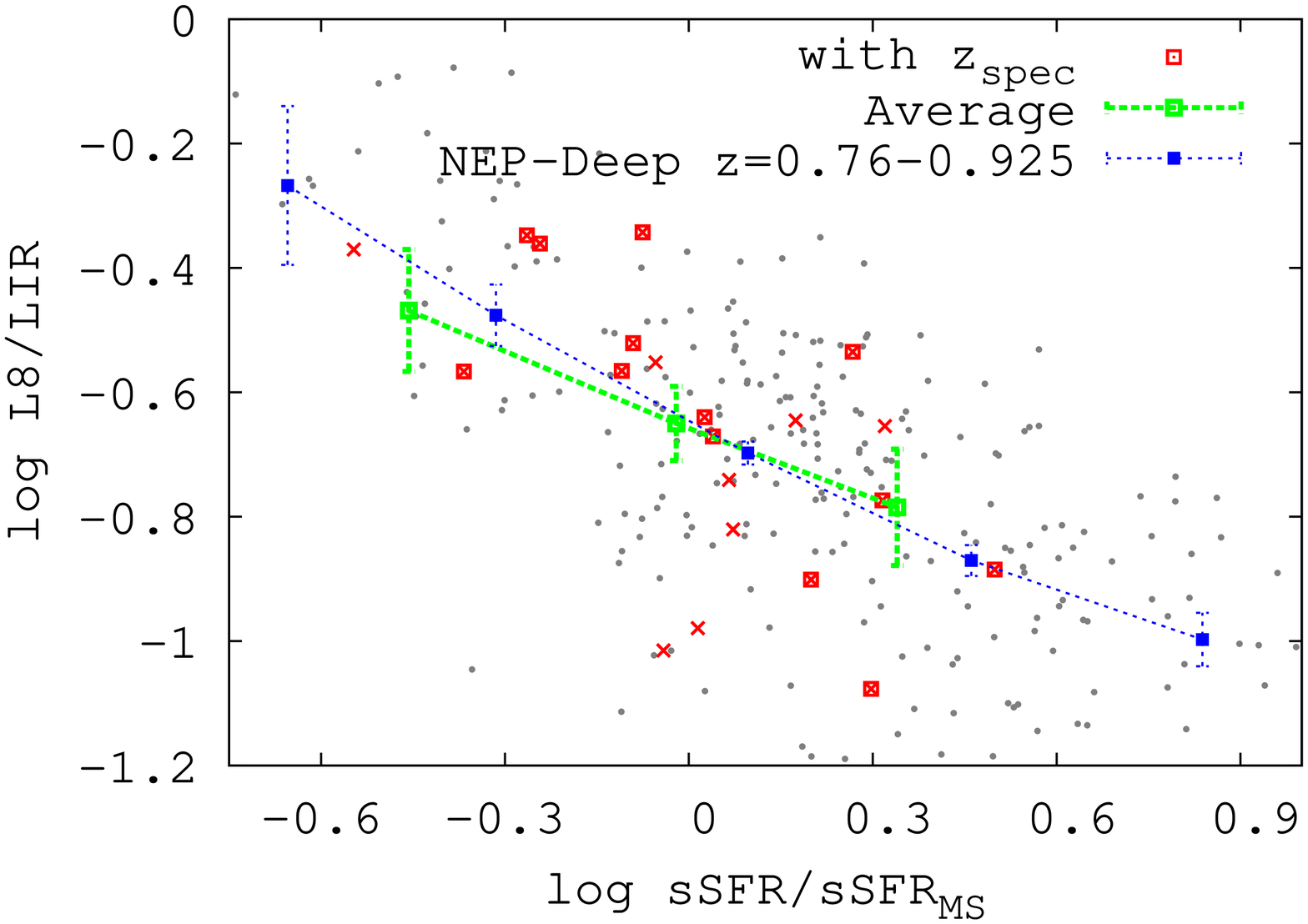}
  \end{minipage}
  \begin{minipage}{0.49\hsize}
    \includegraphics[width=50mm,angle=270]{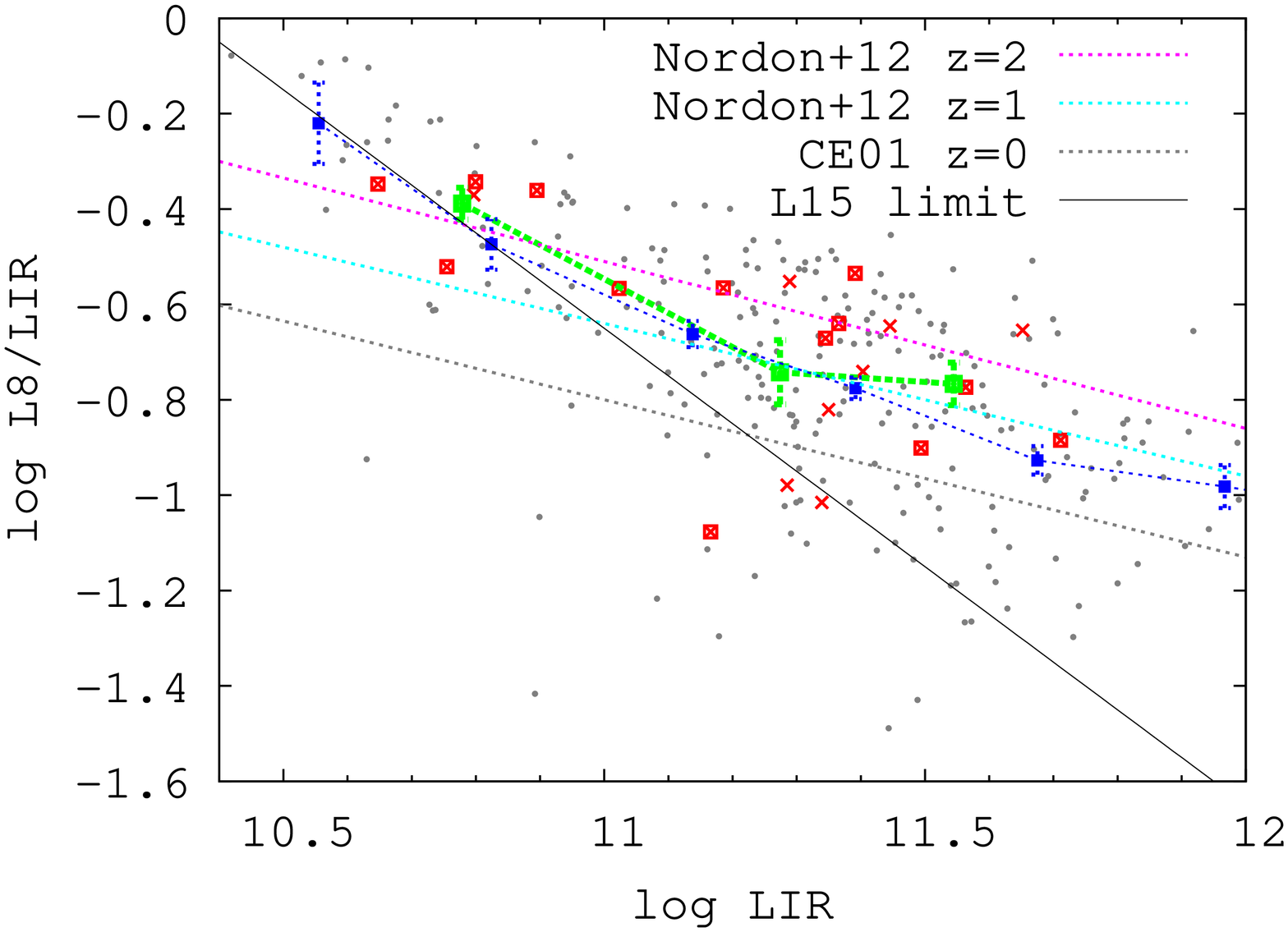}
  \end{minipage}
  \caption{
    Left: L8/LIR vs sSFR divided by that of the main-sequence of the RXJ0152 member galaxies (red crosses) compared with those of the NEP-Deep field galaxies (grey circles).
    The red squares indicate spectroscopically confirmed member galaxies, while red crosses indicate the member galaxies without spectroscopic redshift.
    The green broken line indicates the average of the red points in 0.5 dex bins, where errors are calculated with standard deviation divided by the square root of the number of galaxies in each bin.
    The blue dotted line indicates the average from the NEP-Deep field.
    The redshift range of the NEP-Deep field galaxies is restricted to $z$=0.76-0.925.
    Right: Relation between $L8$/LIR and LIR.
    Symbols are the same as the left panel.
    The grey, cyan, and magenta dotted lines indicate the results from \cite{2001ApJ...556..562C} at $z=$0 and \cite{2012ApJ...745..182N} at $z=1$ and $z=2$, respectively. 
    \label{fig:lml8}}
\end{figure*}

Fig.\ref{fig:lml8} shows the results.
The left panel shows a relation between $L8$/LIR and sSFR normalised by that of main-sequence galaxies.
We applied the sSFR of main-sequence galaxies from \cite{2010A&A...518L..25R} in Fig.\ref{fig:ms}. 
The green line indicates the average value of each point in 0.5 dex size bins, and the errors were calculated with the standard deviation divided by the square root of the number of galaxies in each bin. 
The $L8$/LIR monotonically decreases with sSFR/sSFR$_{MS}$.
Our result is consistent with that from the NEP-Deep field within the errors.
We note that the redshift range of the main-sequence of \cite{2010A&A...518L..25R} is different from ours, which can shift the sSFR/sSFR$_{MS}$ for both the NEP-Deep results and ours, but our conclusion does not change. 
\par
The right panel shows a relation between $L8$/LIR and LIR.
The $L8$/LIR again monotonically decreases with LIR. 
The results from \cite{2001ApJ...556..562C} at $z=0$ and \cite{2012ApJ...745..182N} at $z=1$ and $z=2$ are also shown, which show the redshift evolution of the relation between $L8$/LIR and LIR. 
Our result is consistent with that of the NEP-Deep field. 
Although the slope of the both results is slightly different from the relation from \cite{2012ApJ...745..182N}, they are between the relations at $z=0$ and $z=2$.
At log LIR $<$ 11.2, however, the NEP results and ours are above the relation at $z=2$ in \cite{2012ApJ...745..182N}. This is due to the $L8$ limits, as shown the black solid line in Fig.\ref{fig:lml8}, which leads a lack of low $L8$/LIR galaxies at lower LIR. Hence, at lower luminosity our result can only be compared with the NEP results, whose completeness is similar to our cluster sample.
%  This is because the NEP sample and ours at log LIR $<$ 11.2 is not complete and only higher L8/LIR galaxies are detected. Hence, at lower luminosity our result can only be compared with the NEP results, whose completeness is similar to ours.}
The results of Fig.\ref{fig:lml8} indicate that environment dependence of the redshift evolution of this relation is not remarkable. 
Despite limited sample, all above results support an idea that the behaviour of the $L8$/LIR of cluster galaxies is unaffected by the environment.
%\par
%Fig.\ref{fig:dis_l8lir} gives a hint for understanding the role of the environment in the $L8$/LIR behaviour.

%% file: conc.tex
\section{Discussion and Conclusion}
In this paper, we investigated the behaviour of $L8$/LIR of galaxies in RXJ0152.
Most notably, our study focuses on the $L8$/LIR behaviour, of which the measurement is difficult without using $AKARI$. 
We found that few galaxies in the cluster cores are detected with the L15 band and that galaxies with log $L8$/LIR $<$ -0.75 are located only at the outskirt of the cluster. These results are consistent with previous studies, who found the lack of star-forming galaxies in the cluster core \cite[]{2005ApJ...621..651H}. 
We compared the $L8$/LIR behaviour of the member galaxies with those of field galaxies, obtained from the AKARI NEP-Deep survey, and did not find a clear environmental dependence of the $L8$/LIR behaviour.
The average of $L8$/LIR of both the cluster members and field galaxies decreases with sSFR/sSFR$_{\rm MS}$ and LIR. 
These findings extend previous studies, supporting the idea that the relation between physical parameters of star-forming galaxies is not affected by the environments whereas the fraction of galaxy type is different in different environments.
\par
This idea can be interpreted as that galaxies affected by the environment rapidly evolve into passive galaxies. Some researches support this interpretation. \cite{2013A&A...555A...5N} found that cluster galaxies with peculiar morphology directly evolve into an early type galaxy without having a chance to first evolve into a normal spiral galaxy. \cite{2014ApJ...797..136F} show that galaxies in the cores have formed the bulk of the stars in a short period starburst at earlier epoch while galaxies in the outskirt have various star-formation histories. On the other hand, \cite{2013MNRAS.433..825L} show the importance of galaxy mergers for build up of the stellar mass in brightest cluster galaxies, which are not detected with our L15 band. From these studies, it is implied that galaxies in the cores have experienced a merger, evolved into passive galaxies, and are not detected with the L15 band. 
\par
However, we have some limitations to conclude the above scenario. Most of our sample galaxies are located in the outskirt of the cluster, so that the sample size of the cluster core is small. Furthermore, it is not clear whether the above scenario can apply to other cluster galaxies. Future works should therefore increase the sample size and include galaxies located in cluster cores, by using other galaxy clusters.